\documentclass[,prd,12pt]{revtex4-1}

\usepackage{amsmath}    
\usepackage{graphicx}   
\usepackage{verbatim}   
\usepackage{color}      
\usepackage[caption=false]{subfig} 
\usepackage{hyperref}   
\begin{document}
	
	\title{{\vspace{-14mm}%
			\fontsize{14pt}{10pt}\selectfont
			\textbf{NNLO solution of nonlinear GLR-MQ evolution equation to determine gluon distribution function using Regge like ansatz}
	}}
\author{P. Phukan}
\email{Corresponding author: pragyanp@tezu.ernet.in}
\author{M. Lalung}
\email{Electronic address: mlalung@tezu.ernet.in}
\author{J. K. Sarma}
\email{Electronic address: jks@tezu.ernet.in}
\affiliation{Department of Physics, Tezpur University, Tezpur, Assam-784028, India}
	
	\begin{abstract}
	\noindent \par In this work we have suggested a solution of the Gribov-Levin-Ryskin,Mueller-Qiu (GLR-MQ) nonlinear evolution equation at next-to-next-to-leading order (NNLO). The range of $Q^2$ in which we have solved the GLR-MQ equation is Regge region of the range  6.5 \(\text{GeV}^2\) $\leq $ \(Q^2\) $\leq 25 $ \(\text{GeV}^2\)  and so we have incorporated the Regge like behavior to obtain $Q^2$ evolution of gluon distribution function $G(x, Q^2)$. We have also checked the sensitivity of our results for different values of correlation radius (R) between two interacting gluons, viz.   $R=$ 2 $GeV^{-1}$ and $R=$ 5 $GeV^{-1}$ as well as for different values of Regge intercept $\lambda_G$. Our results are compared with those of most recent  global DGLAP fits obtained by various parametrisation groups viz. PDF4LHC15, NNPDF3.0, HERAPDF15, CT14 and ABM12.
	\end{abstract}
	
	\maketitle
	
	\section{Introduction}
	Parton distribution functions (PDFs) are considered as the most significant tool in hadronic collision processes for the calculation of inclusive cross sections. In perturbative QCD, the scale evolution of the PDFs is well predicted by the Dokshitzer-Gribov-Lipatov-Altarelli-Parisi
	(DGLAP) evolution equation \cite{1,2,3} at large interaction scale \(Q^2\) and the fractional momentum x. At sufficiently  large $Q^2$, the number densities of the partons can
	be evaluated  by solving the DGLAP equations from which the emission of partons during the process can be spotted and then over a broad range of x and \(Q^2\), a comparison with the data is performed. Further for obtaining a good global fit to the data, the initial distributions
	are iterated. The initial distributions are the non-perturbative inputs, that perturbative QCD cannot
	predict.
	\par The data from lepton-proton deep inelastic scattering (DIS), particularly the DIS data from ep collisions at DESY-HERA play a key role in these analysis especially in the region of small-x. It is evident from the data that there is a sharp growth of the gluon density towards small-x \cite{1,2}. This is also well predicted by the solutions of linear DGLAP equation. However, the gluon density cannot grow forever because hadronic cross-sections comply with	the unitary bound known as Froissart Bound \cite{4}. For this purpose a distinguishable effect known as gluon recombination which is supposed to be responsible for the mechanism that unitarize the cross section at high energies or at small-x. In other words, at small-x the number of gluon	will be so large that they will spatially overlap, resulting recombination of gluons. But in the derivation of the DGLAP equation, the gluon-gluon interaction terms are overlooked. Thus a modification in the linear DGLAP equation is required in order to take care of the nonlinear corrections due to gluon recombination. 
	\par
	The the H1 Collaboration \cite{5} at HERA has able to calculate the proton structure function \(F_2\left(x,Q^2\right)\) down to \(x\sim 10^{-5}\) though it is in the perturbative region. These data have been included in the recent global analyses by the CTEQ \cite{7} and  MRST \cite{6} collaborations. Though DGLAP equation comply with the experimental data quite accurately in a wide range of x and \(Q^2\), it fails to provide a favorable explanation in fitting the H1 collaboration data towards the region of large \(Q^2\left(>4\text{GeV}^{-2}\right)\) and in the region of small \(Q^2\) (1.5 \(\text{GeV}^{-2}<Q^2<4
	\text{GeV}^{-2}\)) \cite{8,9} simultaneously. Furthermore, in the NLO treatment of MRST2001 \cite{36} when both these regions were taken into consider, a good fit was obtained but a negative gluon distribution was encountered. Likewise in the NLO set CTEQ6M \cite{37},the problem of negative gluon distribution also appears. This implies that towards smaller values of x and \(Q^2\), constraining that \(Q^2\)$\geq $ \(\Lambda ^2\), $\Lambda $ being the QCD cutoff parameter, it is possible to observe gluon recombination effects which lead to nonlinear power corrections to the linear DGLAP equation.
	\par
	Gribov, Levin and Ryskin in the ref.\cite{10} and Mueller and Qiu in the ref.\cite{11} have calculated the nonlinear terms and they formulated these shadowing corrections to obtain a	new evolution equation commonly known as GLR-MQ equation. This equation deals with a new quantity \(G^2\left(x,Q^2\right)\) interpreted as the two gluon distribution function per unit area of the hadron. In addition to the explanation of gluon saturation phenomena, GLR-MQ equation predicts a critical line which supposed to separate the gluon saturation regime and the perturbative regime valid in this critical line border\cite{12}. Most significantly GLR-MQ equation introduces a characteristic momentum scale \(Q_s^2\) which provides the measure of the density of the saturated gluons.
	\par
	The GLR-MQ equation is regarded as a hypothetical link between perturbative and non perturbative region. There has been some work in recent years inspired by
	GLR-MQ approach \cite{12,13}. The solution of GLR-MQ equation provides the determination of the saturation momentum that incorporates physics in addition to that	of the linear evolution equations commonly used to fit DIS data. In our previous works we obtained a solution of the \(Q^2\) dependence of gluon	distribution from GLR-MQ in leading order (LO) \cite{14}  as well as next-to-next-leading order (NLO) \cite{15}. In the present work, we adopt the Regge-like parametrizations
	to obtain a solution of the nonlinear GLR-MQ equation up to next-to-next-leading order(NNLO) and a direct comparison of our results with those of the global DGLAP fits obtained by various collaborations viz. NNPDF3.0 \cite{29}, HERAPDF1.5 \cite{30}, CT14 \cite{31}, ABM12 \cite{33} and PDF4LHC \cite{32}.	
	\section{Theory}
	GLR-MQ equation deals with the number of partons increased through gluon splitting as well as the number of partons decreased through gluon recombination in a phase space cell ($\Delta \text{ln}(1/x)\Delta \text{ln}{Q^2}$). Therefore the balance equation for emission and recombination of partons can be formulated as \cite{3,6}
	\begin{equation*}
	\frac{\partial \rho\left(x, Q^2\right)}{\partial \text{ln(1/x)} \text{lnQ}^2}= \frac{\alpha_s(Q^2)N_c}{\pi}\rho(x,Q^2)-\frac{{\alpha_s}^2(Q^2)\gamma}{Q^2}[\rho(x,Q^2)]^2.
	\end{equation*}
	Here $\rho=\frac{xg(x,Q^2)}{\pi R^2}$, where R is the correlation radius between two interacting gluons, $\pi R^2$ is the target area. The factor $\gamma$ is evaluated by Muller and Quie that is found to be $\gamma=\frac{81}{16}$ for $N_c=3$\cite{6}. Now in terms of gluon distribution function \(G\left(x,Q^2\right.\))= \(\text{xg}\left(x,Q^2\right.\)) the GLR-MQ equation can be written in standard form \cite{16}
	\begin{equation}
	\frac{\partial G\left(x, Q^2\right)}{\partial \text{lnQ}^2}= \frac{\partial G\left(x, Q^2\right)}{\partial \text{lnQ}^2}| _{\text{DGLAP}}-\frac{81}{16}\frac{\alpha
		_s{}^2\left(Q^2\right)}{R^2Q^2}\int _x^1\frac{\text{d$\omega $}}{\omega }G^2\left(\frac{x}{\omega },Q^2\right).
	\end{equation}
	The first term of the RHS in eq.(1) represents the double-leading logarithmic approximation (DLLA) linear DGLAP term while the second term is the shadowing correction due to the nonlinearity
	in gluon density. At small-x region, the contribution of quark-gluon diagrams are very negligible. For the correlation radius, \(R=R_H\), shadowing correction is negligibly
	small whereas for \(R<<R_H\), shadowing correction is expected to be large, where \(R_H\) is the radius of the hadron \cite{17,18}.
	\par
	We introduce a variable \(t=\ln \left(\frac{Q^2}{\Lambda ^2}\right),\) where $\Lambda $ is the QCD cutoff parameter. Now Considering the terms up
	to NNLO, \(\alpha _s\)(t) takes the following form
	\begin{equation}
	\alpha _s(t)= \frac{4\pi }{\beta _0 t^2}\left\{t-b \text{lnt}+b^2\left( \ln ^2t-\text{lnt}-1\right)+c\right\},
	\end{equation}
	where \(b=\frac{\beta _1}{\beta _2{}^2},c=\frac{\beta _2}{\beta _0{}^3}, \beta _0=11-\frac{2}{3}N_f,\beta _1=102-\frac{38}{3}N_f.\)
	Here we consider the number of color charges, \(N_c\) and the number of quark flavors, \(N_f\) as 3 and 4 respectively. Now in terms of
	the variable t, eq.(1) can be expressed as
	\begin{equation}
	\frac{\partial G(x, t)}{\partial t }= \frac{\partial G(x, t)}{\partial t }| _{\text{DGLAP}}-\frac{81}{16}\frac{\alpha _s{}^2(t)}{R^2\Lambda ^2e^t}\int
	_x^1\frac{\text{d$\omega $}}{\omega }G^2\left(\frac{x}{\omega },t\right).\end{equation}
	\par Ignoring the quark contribution to the gluon rich distribution function, we can write the first term of the eq.(3) of the form
	\begin{equation}\frac{\partial G(x, t)}{\partial t }| _{\text{DGLAP}}=\int _x^1P_{\text{gg}}(\omega )G\left(\frac{x}{\omega },t\right)\text{d$\omega $}.\end{equation}
	Considering up to NNLO terms, the splitting function \(P_{\text{gg}}(\omega )\) can be expanded as powers of \(\alpha _s(t)\),
	\begin{equation}
	P_{\text{gg}}(\omega )=\frac{\alpha _s(t)}{2\pi }P_{\text{gg}}^0(\omega )+\left(\frac{\alpha _s(t)}{2\pi }\right)^2P_{\text{gg}}^2(\omega )+\left(\frac{\alpha
		_s(t)}{2\pi }\right)^3P_{\text{gg}}^3(\omega ).
	\end{equation}
	The corresponding splitting functions involved in eq.(5) are\\ 
	LO splitting function \cite{19}
	\begin{equation*}
	P_{\text{gg}}{}^0(\omega )= 6\left(\frac{1-\omega }{\omega }+\frac{\omega }{(1-\omega )_+}+\omega (1-\omega )\right)+\left(\frac{11}{2}-\frac{2N_f}{3}\right)\delta
	(1-\omega ),
	\end{equation*}
	NLO splitting function
	\begin{align*}
	P_{gg}^1=&C_F T_f \left\{-16+8 \omega +\frac{20 \omega ^2}{3}+\frac{4}{3 \omega }-(6+10 \omega ) {ln\omega}-2 (1+\omega ) \ln
	^2 \omega \right\}\\
	&+N_c T_f \left\{2-2 \omega +\frac{26}{9} \left(\omega ^2-\frac{1}{\omega }\right)-\frac{4}{3} (1+\omega ) \text{ln$\omega $}-\frac{20}{9}p(\omega
	)\right\}
	\\
	&+N_c^2 \bigg\{\frac{27 (1-\omega )}{2}+\frac{67}{9} \left(\omega ^2-\frac{1}{\omega }\right)-\left(\frac{25}{3}-\frac{11 \omega }{3}+\frac{44 \omega
		^2}{3}\right) \text{ln$\omega $}\\
	&+4(1+\omega ) \ln ^2 \omega +\left(\frac{67}{9}+\ln ^2 \omega -\frac{\pi ^2}{3}\right) p(\omega )\\
	&-4\text{ln$\omega$} \ln  (1-\omega ) p(\omega )+2 p(-\omega ) S_2(\omega )\bigg\},
	\end{align*} where \hspace{0.5cm}
 $p(\omega )=\frac{1}{1-\omega }+\frac{1}{\omega }-2+\omega (1-\omega )$,\\
 
	\hspace{1cm} $S_2(\omega )=\int _{\frac{\omega }{1+w}}^{\frac{1}{1+\omega }}\frac{\text{dz}}{z} \ln \left(\frac{1-z}{z}\right) \underset{\omega }{\overset{\text{small}}{\longrightarrow
	} }\frac{1}{2}\ln ^2\omega -\frac{\pi ^2}{6}+O(\omega )$,\\ 

	\hspace{1.4cm} $C_F=\frac{N_c^2-1}{2N_c}, T_f=\frac{1}{2}N_f$,\\ and NNLO splitting function \cite{20}
	\begin{align*}
	P_{\text{gg}}^2=&2643.52 D_0+4425.89 \delta  (1-\omega )+3589 L_1-20852+3968 \omega -3363 \omega ^2\\
	&+4848 \omega ^3+L_0 L_1 \left(7305+8757 L_0\right)+274.4
	L_0-7471 L_0^2+72 L_0^3-144 L_0^4\\ &+\frac{14214}{\omega }+\frac{2675.8 L_0}{\omega }+N_f \bigg\{-412.172 D_0-528.723 \delta  (1-\omega )-320 L_1\\ &-350.2\,
	+755.7 \omega -713.8 \omega ^2+559.3 \omega ^3+L_0 L_1 \left(26.15\, -808.7 L_0\right)+1541 L_0\\ &+491.3 L_0^2+\frac{832 L_0^3}{9}+\frac{512 L_0^4}{27}+\frac{182.96}{\omega
	}+\frac{157.27 L_0}{\omega }\bigg\}
\end{align*}
\begin{align*}
&+N_f^2 \bigg\{-\frac{16 D_0}{9}+6.463 \delta  (1-\omega )-13.878\, +153.4 \omega -187.7 \omega ^2\\ &+52.75 \omega
	^3-L_0 L_1 \left(115.6\, -85.25 \omega +63.23 L_0\right)-3.422 L_0\\ &+9.68 L_0^2-\frac{32 L_0^3}{27}-\frac{680}{243 \omega }\bigg\},
	\end{align*}
	where $D_0=\frac{1}{(1-\omega )_+}, L_0=\ln\omega$ and $L_1= \ln (1-\omega )$.
	\par Now considering all these terms, the DGLAP equation takes up the following form in NNLO\\
	\begin{equation}
	\begin{split}
	\frac{\partial G(x, t)}{\partial t }| _{\text{DGLAP}}=&\frac{3\alpha _s(t)}{\pi }\bigg[\bigg\{\frac{11}{12}-\frac{N_f}{18}+\ln (1-x)\bigg\}G(x,t)\\ &+\int
	_x^1\text{d$\omega $}\bigg\{\frac{\text{$\omega $G}\bigg(\frac{x}{\omega },t\bigg)-G(x,t)}{(1-\omega )}\\ &+\bigg(\omega (1-\omega )+\frac{1-\omega }{\omega
	}\bigg)G\bigg(\frac{x}{\omega },t\bigg)\bigg\}\bigg]\\
	&+\bigg(\frac{\alpha _s(t)}{2\pi }\bigg)^2I_1^g(x,t)+\bigg(\frac{\alpha _s(t)}{2\pi }\bigg)^3I_2^g(x,t),
	\end{split}
	\end{equation} where $I_1^g(x,t)=\int _x^1\text{d$\omega $}\left[P_{\text{gg}}^1(\omega )G\left(\frac{x}{\omega },t\right)\right]
	\text{and{ }} I_2^g(x,t)=\int _x^1\text{d$\omega $}\left[P_{\text{gg}}^2(\omega )G\left(\frac{x}{\omega },t\right)\right]$.
	\par
	For simplicity in our calculations, we consider two numerical parameters \(T_0\) and \(T_1\) such that \(T^2(t)\) = \(T_0T(t)\) and \(T^3(t)=T_1T(t)\),
	where \(T(t)=\alpha _s/2\pi\). \(T_{0 }\) and \(T_1\) are not arbitrary parameters. These numerical parameters are determined by phenomenological analysis. These  are
	obtained from the particular range of \(Q^2\) under our study and by a suitable choice of \(T_0\) and \(T_1\) we
	can reduce the difference between \(T^2\)(t) and \(T_0T(t)\) as well as \(T^3\)(t) and \(T_1T(t)\) to minimum such that the consideration of the parameters \(T_0\) and \(T_1\) doesn't give any abrupt
	change in our work.\par
\begin{figure}[h]
	\centering
	\subfloat{%
		\includegraphics[clip,width=0.38\columnwidth]{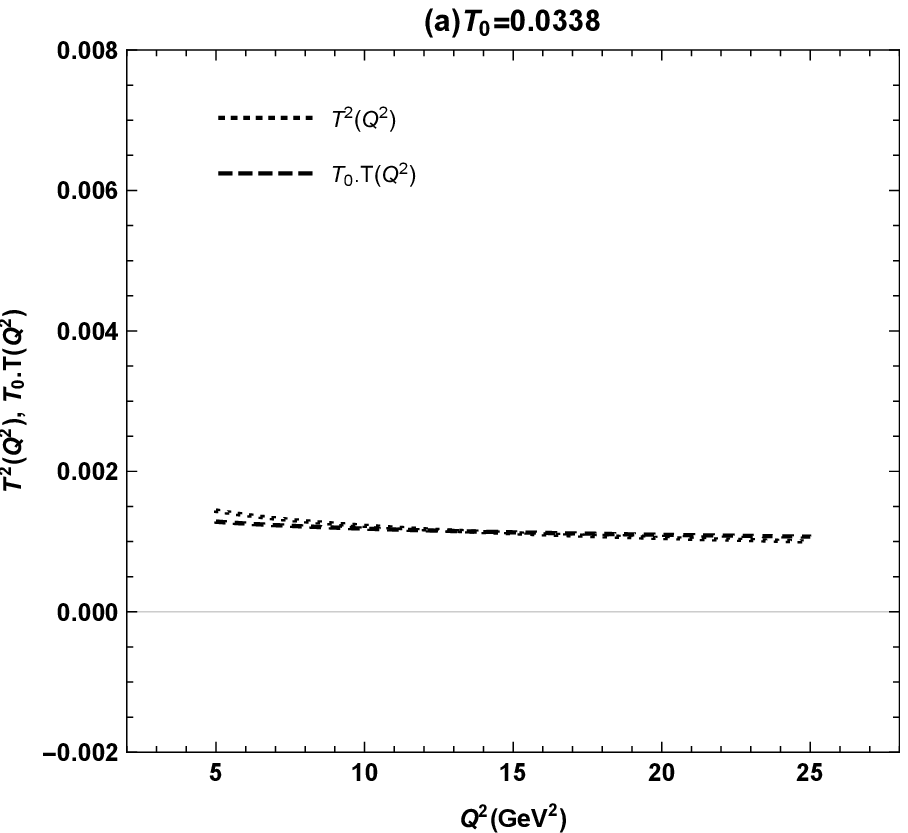}}
	\hspace{5mm}
	\subfloat{%
		\includegraphics[clip,width=0.38\columnwidth]{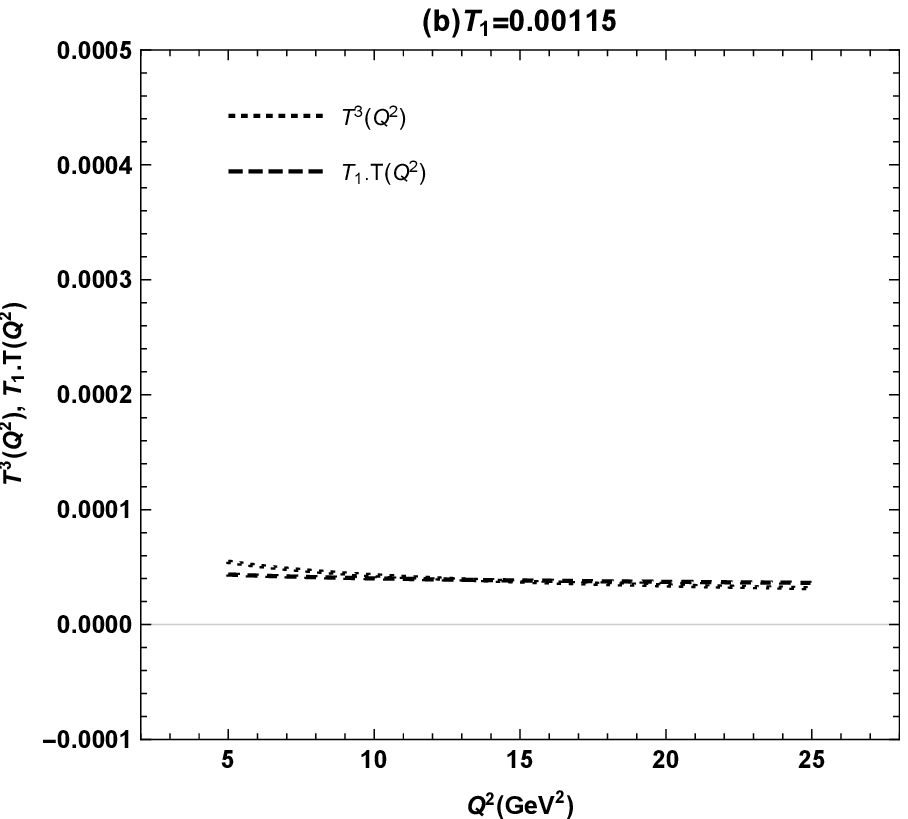}}
	\caption{Comparison of $T^2(t)$ and $T_0.T(t)$ as well as $T^3(t)$ and $T_1.T(t)$ vs $Q^2$}
	
\end{figure}

	To get an analytical solution of the GLR-MQ equation we incorporate a Regge-like behavior of gluon distribution function. The behavior of structure functions at small-x is well explained in terms of Regge-like ansatz \cite{21}. For small-x, the Regge behavior of the sea	quark and antiquarks distribution is given by \(q_{\text{sea}}\sim x^{-\alpha _P}\) corresponding to a pomeron exchange with an intercept of \(\alpha _P=1\). But the valence-quark distribution for small-x given by \(q_{\text{val}}(x)\sim x^{-\alpha _R}\) corresponding to a reggeon exchange with an intercept of \(\alpha _R=0.5.\) At moderate \(Q^2\), the leading order calculations in \(\text{ln}(1/x)\) with fixed
	value of \(\alpha _s\) predicts a steep power law behavior of \(\text{xg}\left(x,Q^2\right)\sim x^{-\lambda _G}\), where \(\lambda _G=\left(4\alpha
	_s\left.N_c\right/\pi \right)\text{ln2}\) $\approx $0.5 for \(\alpha _s=0.2\), as appropriate for \(Q^2>4\text{GeV}^2\) \cite{22,23,24}.
	
	\par
	
	To determine the gluon distribution function we try to solve GLR-MQ equation by  considering a simple form of Regge like behavior
	given as
	\begin{equation}
	G(x,t)=H(t)x^{-\lambda _G},\end{equation}
	which implies \begin{equation}G\left(\frac{x}{\omega },t\right)=H(t)x^{-\lambda _G}\omega ^{\lambda _G}=G(x,t)\omega ^{\lambda }\end{equation}
	and \begin{equation}G^2\left(\frac{x}{\omega },t\right)=\left(H(t)x^{-\lambda _G}\right){}^2\omega ^{2\lambda _G}=G^2(x,t)\omega ^{2\lambda _G},\end{equation}\\
	where \(\lambda _G\) is the Regge intercept for gluon distribution function while  H(t) is a function of t. Several literatures ref. \cite{25,27}, deal with this form of Regge like behavior. In accordance with the Regge theory, at small-x, both gluons and sea quarks behaviors are controlled by the same singularity factor
	in the complex angular momentum plane \cite{21}. At small x, since the Regge intercepts, \(\lambda _G\) of all spin-independent singlet, non-singlet and gluon structure
	functions should tend to 0.5 \cite{28}, it is also expected that at \(\lambda _G\approx 0.5\), our theoretical results comply with the experimental
	data and parametrization .\par
	Substituting eqs. (7), (8) and (9) in eq.(3) the GLR-MQ equation becomes
	\begin{equation}
	\frac{t^2 }{t-a \text{lnt}+b^2 \text{lnt}^2-b^2+c}\frac{\partial G(x, t)}{\partial t }=\rho (x)G(x,t)-\frac{\phi (x)G^2(x,t)}{e^t},
	\end{equation}
	where\begin{align*}
	\rho  (x) &=3 A_f \left(\frac{11}{12}-\frac{N_f}{18}+\ln (1-x)+\frac{2}{2+\lambda _G}\right)-3 A_f \left(\frac{2 x^{\lambda _G+2}}{\lambda _G+2}+\frac{x^{\lambda
			_G}}{\lambda _G}-\frac{1}{\lambda _G}-x+1\right)\\ &+ \frac{T_0}{2}A_f\int _x^1\text{d$\omega $P}_{\text{gg}}^1(\omega )\omega ^{\lambda _G}+\frac{T_1}{2}
	A_f\int _x^1\text{d$\omega $P}_{\text{gg}}^2(\omega)\omega ^{\lambda _G},\\
	\phi (x)&=T_0A_f\frac{81}{16}\frac{2\pi ^2}{R^2\Lambda ^2}\left(\frac{1-x^{\lambda _G}}{2\lambda _G}\right),\text{ }
	A_f=\frac{4}{\beta _0},\text{ }\\ a&=b+b^2, \text{  } b=\beta_1/{\beta_0}^2 \text{ } \text{and} \text{ } c=\beta_2/{\beta_0}^3.\end{align*}
	eq.(10) is a partial differential equation, the solution of which is of the form\\
	\begin{equation}
	G(x,t)=\frac{e^{\left(\frac{a }{t}-\frac{ b^2}{t}-\frac{c}{t}-\frac{b^2\ln ^2t}{t}\right)\rho (x)} t^{\left(1+\frac{a }{t}-\frac{2 b^2}{t}\right)\rho
			(x)}}{\int _1^t\text{dz}\frac{1}{z^2}\phi (x)e^{\Delta (x,z)} \zeta(x,z)+C} \text{  }\text{,} \end{equation}
	where \begin{align*}\Delta (x,z)=&\left(\frac{a }{z}-\frac{b^2}{z}-\frac{ c}{z}+\text{lnz}+\frac{a \text{lnz}}{z}-\frac{2\text{  }b^2 \text{lnz}}{z}-\frac{
		b^2 \ln ^2 z}{z}\right)\rho (x)-z \text{  },\\
	\zeta(x,z)=&-b^2+c+z-a \text{lnz}+b^2 \ln ^2z
	\end{align*} 
 and $\rho(x)$, $\phi(x)$ are defined earlier in eq.(10)
	and C is a constant to be determined using initial conditions of the gluon distributions for a given \(t_0\), where \(t_0=\text{ln}\left(\frac{Q_0^2}{\Lambda
		^2}\right)\),
	\begin{equation}
	G\left(x,t_0\right)=\frac{e^{\left(\frac{a }{t_0}-\frac{ b^2}{t_0}-\frac{c}{t_0}-\frac{b^2\ln ^2t_0}{t_0}\right)\rho (x)} t_0{}^{\left(1+\frac{a
			}{t_0}-\frac{2 b^2}{t_0}\right)\rho (x)}}{\int _1^{t_0}\text{dz}\frac{1}{z^2}\phi (x)e^{\Delta (x,z)} \zeta(x,z)+C}\text{  },\end{equation}
	which implies\begin{equation} C=\frac{ e^{\left(\frac{a }{t_0}-\frac{ b^2}{t_0}-\frac{c}{t_0}-\frac{b^2\ln ^2t_0}{t_0}\right)\rho (x)} t_0^{\left(1+\frac{a }{t_0}-\frac{2 b^2}{t_0}\right)\rho
			(x)}-\int _1^{t_0}\text{dz}\frac{1}{z^2}\phi (x)e^{\Delta (x,z)} \zeta(x,z)G\left(x,t_0\right)}{G\left(x,t_0\right)}.\end{equation}
	\par Now substituting C from eq.(13) 
	we obtain the t (or \(Q^2\)) evolution of gluon distribution function G(x,t) for fixed x in NNLO as
	\begin{equation}
	\begin{split}
	G(x,t)=\frac{G(x,t_0)e^{\left(\frac{a }{t}-\frac{ b^2}{t}-\frac{c}{t}-\frac{b^2\ln ^2t}{t}\right)\rho (x)} t^{\left(1+\frac{a }{t}-\frac{2 b^2}{t}\right)\rho
			(x)}}{G(x,t_0)\int _{t_0}^t\text{dz}\frac{1}{z^2}\phi (x)e^{\Delta (x,z)} \zeta(x,z)+e^{\left(\frac{a }{t_0}-\frac{
				b^2}{t_0}-\frac{c}{t_0}-\frac{b^2 \ln ^2t_0}{t_0}\right)\rho (x)} t_0{}^{\left(1+\frac{a }{t_0}-\frac{2 b^2}{t_0}\right)\rho (x)}} .
	\end{split}
	\end{equation}
	Thus by solving GLR-MQ equation semi numerically, we have obtained an expression for the $Q^2$ or $t$ evolution of gluon distribution function $G(x,t)$ up to NNLO. From this final expression we can easily anticipate the t-evolution of \(G\left(x,Q^2\right)\) for a particular value of x by choosing a suitable
	input. 
	\section{Results}
	In this work we have obtained the solution of nonlinear GLR-MQ evolution equation to determine the t (or \(Q^2\)) evolution of gluon distribution function
	\(G\left(x,Q^2\right)\) up to NNLO. Also we have compared our results with those recent global DGLAP fits obtained by various collaborations viz. NNPDF3.0 \cite{29}, HERAPDF1.5 \cite{30}, CT14 \cite{31}, ABM12 \cite{33} and PDF4LHC \cite{32}.
		\begin{figure}[b]
		\centering
		\subfloat{%
			\includegraphics[clip,width=0.38\columnwidth]{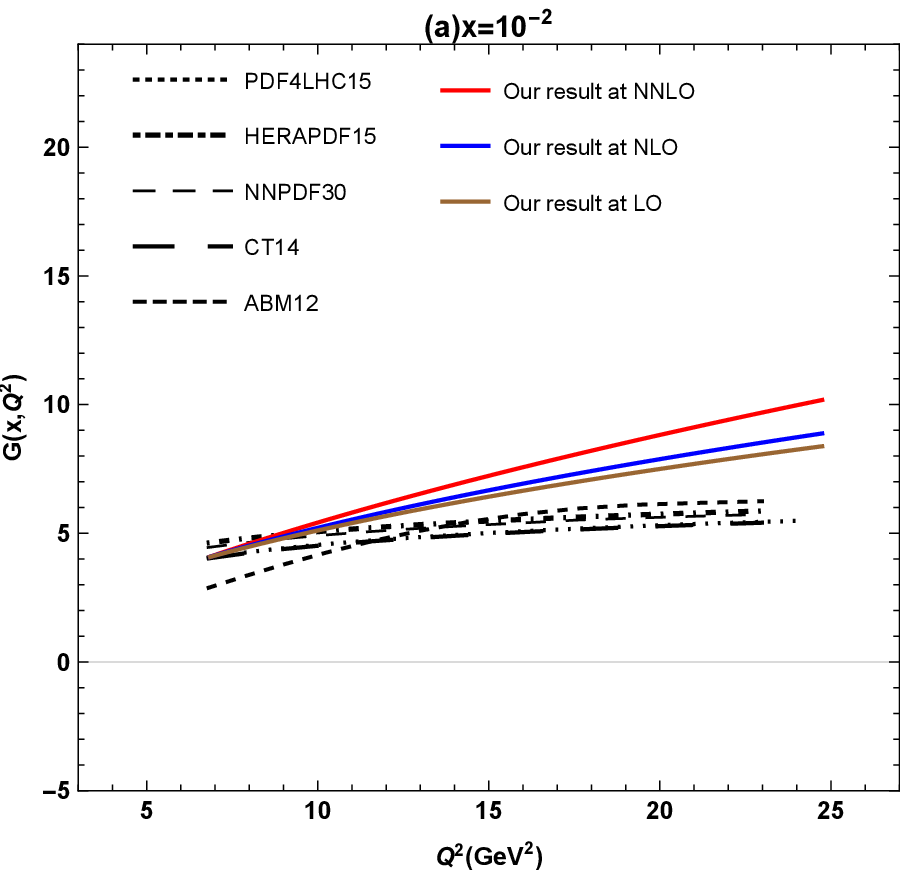}}
		\hspace{5mm}
		\subfloat{%
			\includegraphics[clip,width=0.38\columnwidth]{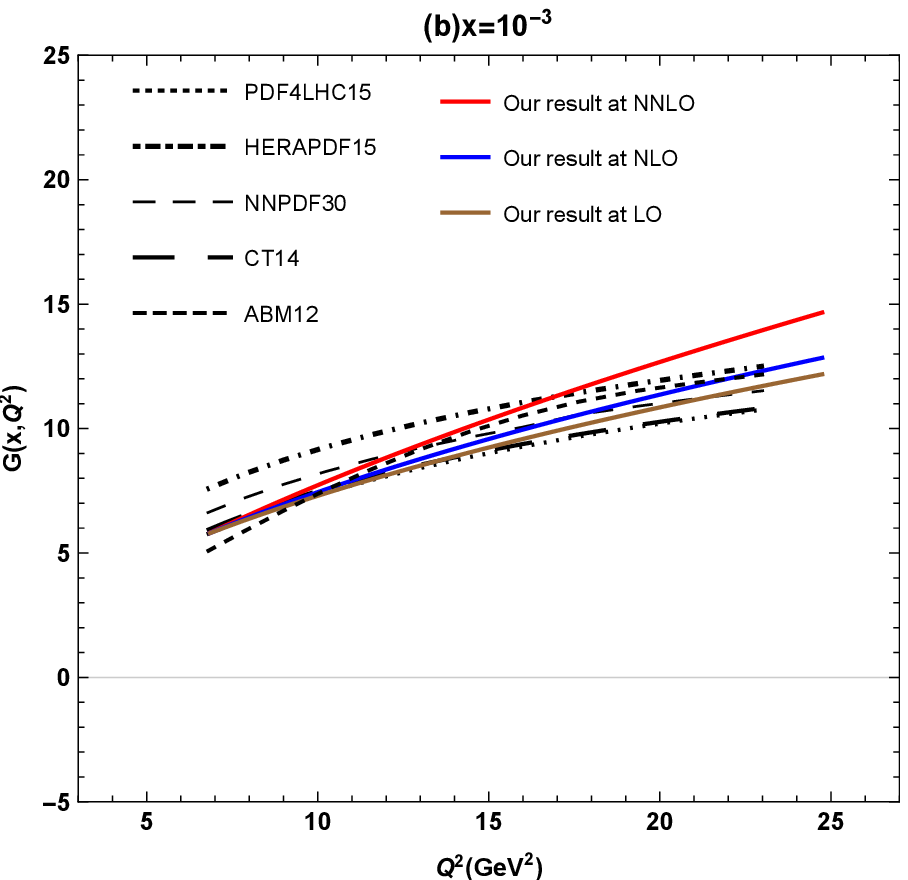}}
		\vspace{1mm}
		\subfloat{%
			\includegraphics[clip,width=0.38\columnwidth]{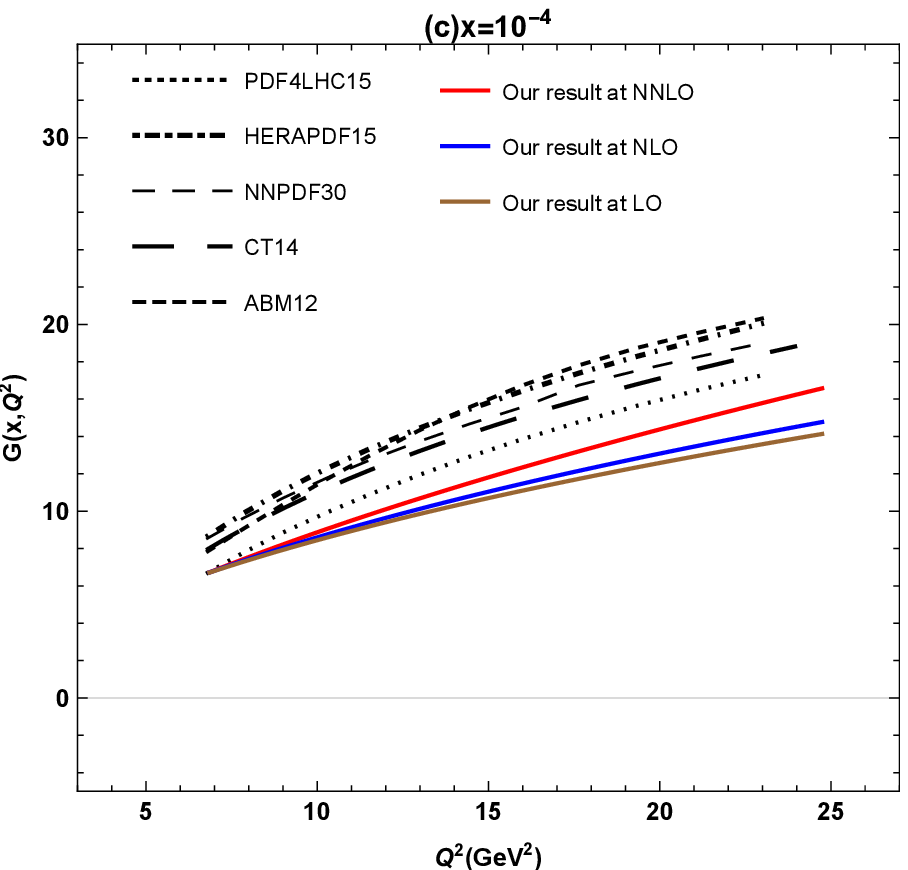}}
		\hspace{5mm}
		\subfloat{%
			\includegraphics[clip,width=0.38\columnwidth]{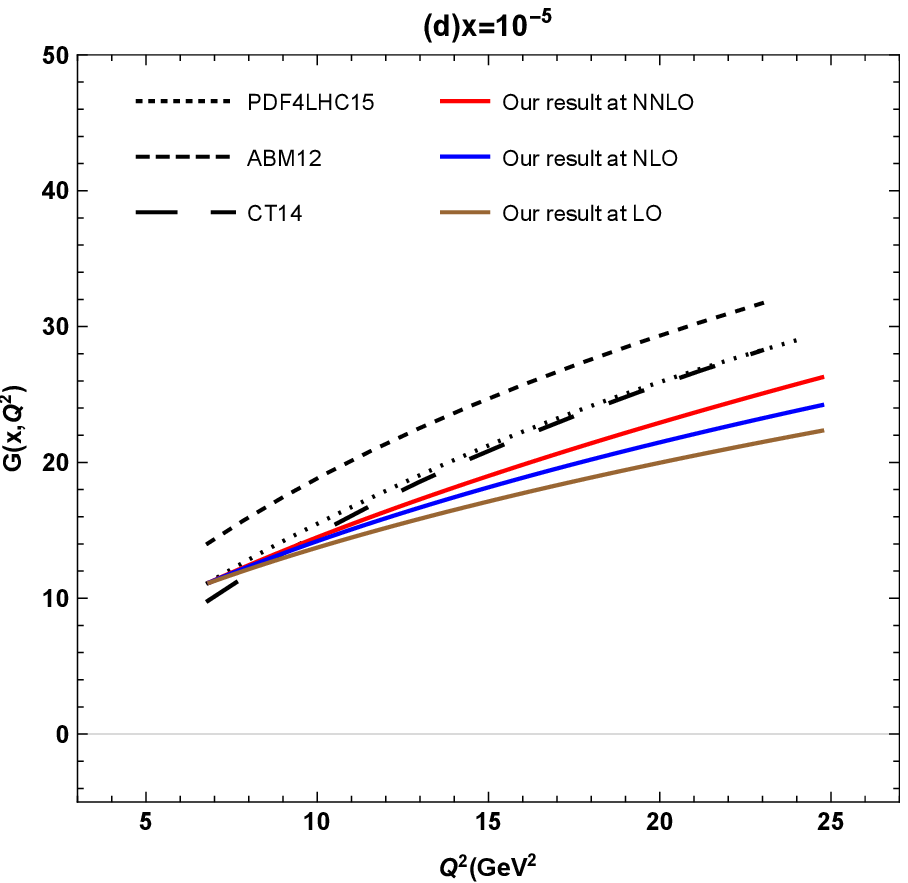}}
		
		\caption{$Q^2$ evolution of $G(x,Q^2)$ for $R=2 GeV^{-1}$. \textit{Solid red lines} are our NNLO result while \textit{solid blue lines} and \textit{solid brown lines} are our NLO and LO results respectively. \textit{Dotted lines} are from the PDF4LHC15 set, \textit{dot dashed lines} are from HERAPDF15 set, \textit{dashed lines} are from NNPDF30 and \textit{absolute dashed lines} are from CT14.}
		
	\end{figure} The NNPDF3.0 set uses a global dataset which includes various HERA data as well as relevant LHC data.
	The QCD fit analysis of the combined HERA-I inclusive deep inelastic cross-sections have been extended to include combined HERA-II measurement
	at high $Q^2$ resulting into HERAPDF1.5 sets. The CT14 includes the data from LHC experiments as well as the new D$\emptyset $ charged lepton asymmetry data. The ABM12 set results from the global analysis of DIS and hadron collider data including the available LHC data for standard candle processes such as \(W^{\pm }\) and Z-boson and \(\overset{-}{t}t\) production. The PDF4LHC15 set is the updated recommendation of PDF4LHC	group for the usage of sets of PDFs and the assessment of PDF and \(\text{PDF}+\alpha _s\) uncertainties which is suitable for applications at the
	LHC Run II.
	\begin{figure}[h]
		\centering
		\subfloat{%
			\includegraphics[clip,width=0.38\columnwidth]{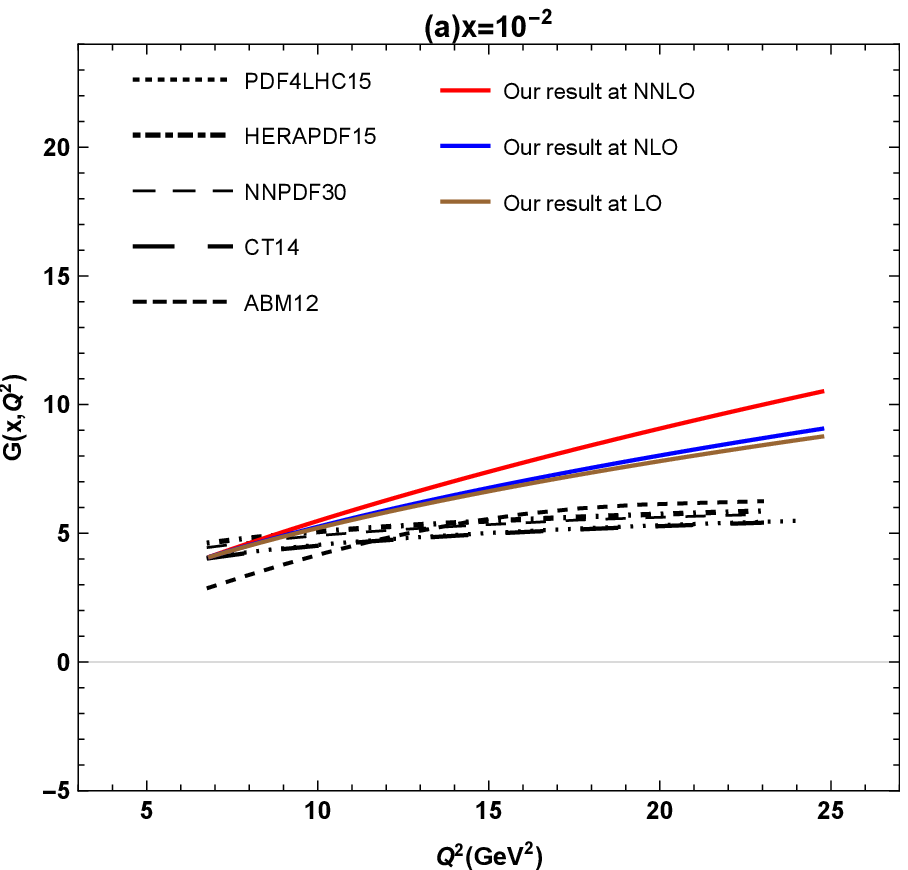}}
		\hspace{5mm}
		\subfloat{%
			\includegraphics[clip,width=0.38\columnwidth]{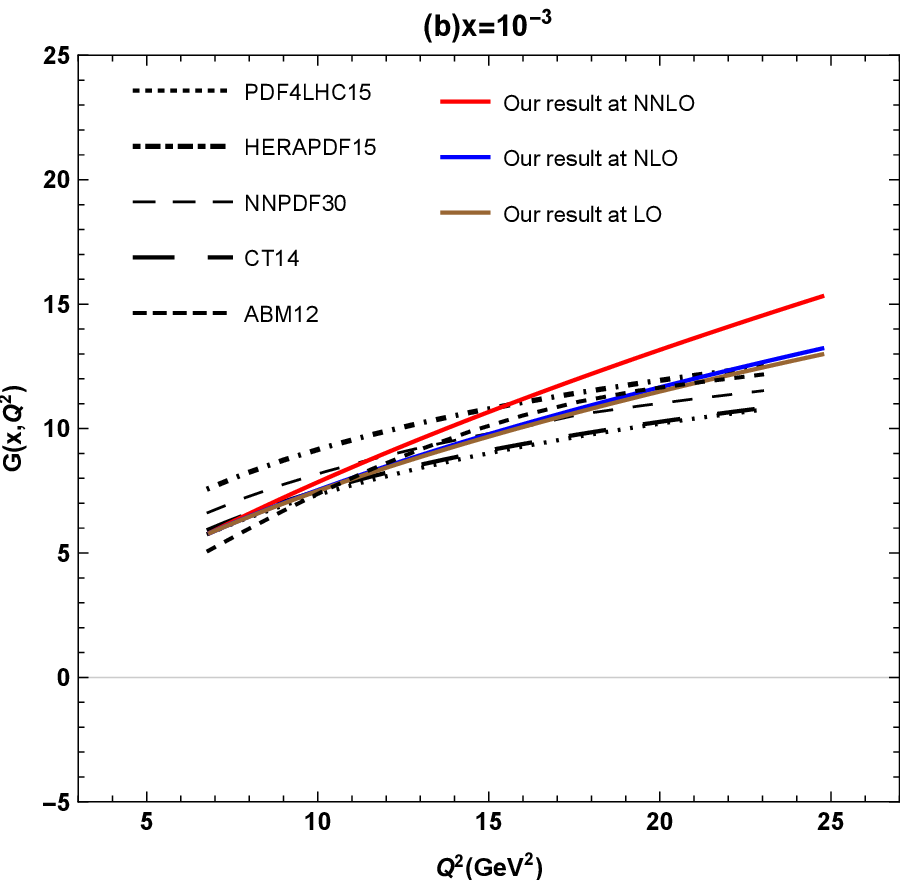}}
		\vspace{1mm}
		\subfloat{%
			\includegraphics[clip,width=0.38\columnwidth]{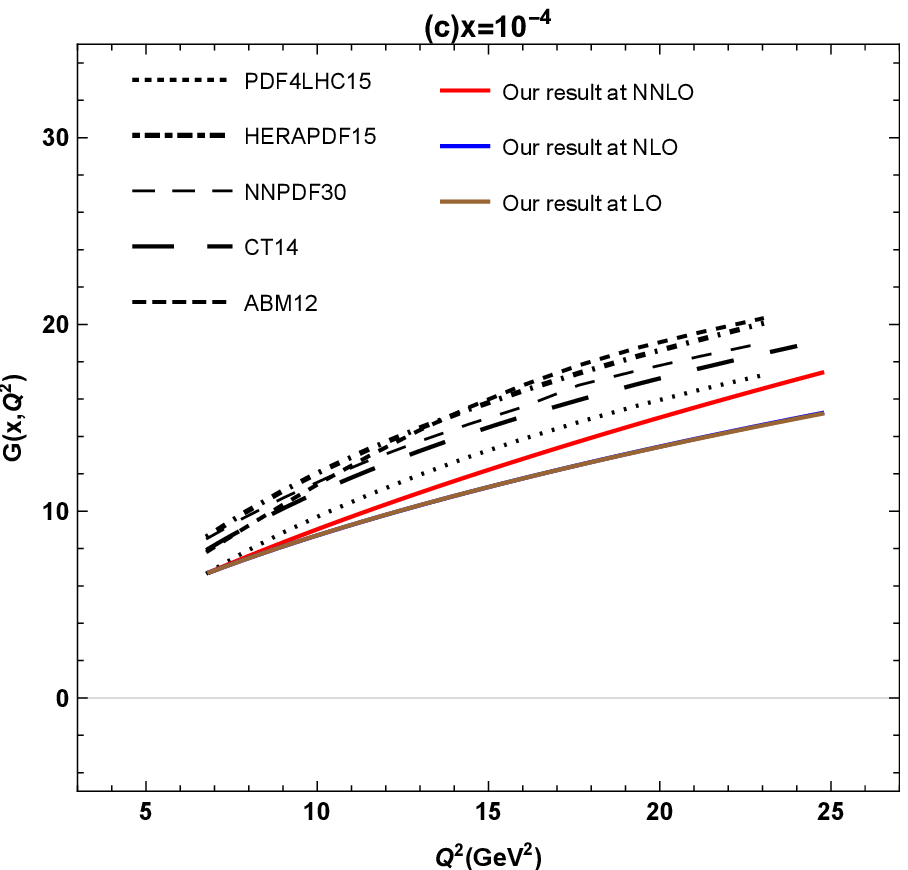}}
		\hspace{5mm}
		\subfloat{%
			\includegraphics[clip,width=0.38\columnwidth]{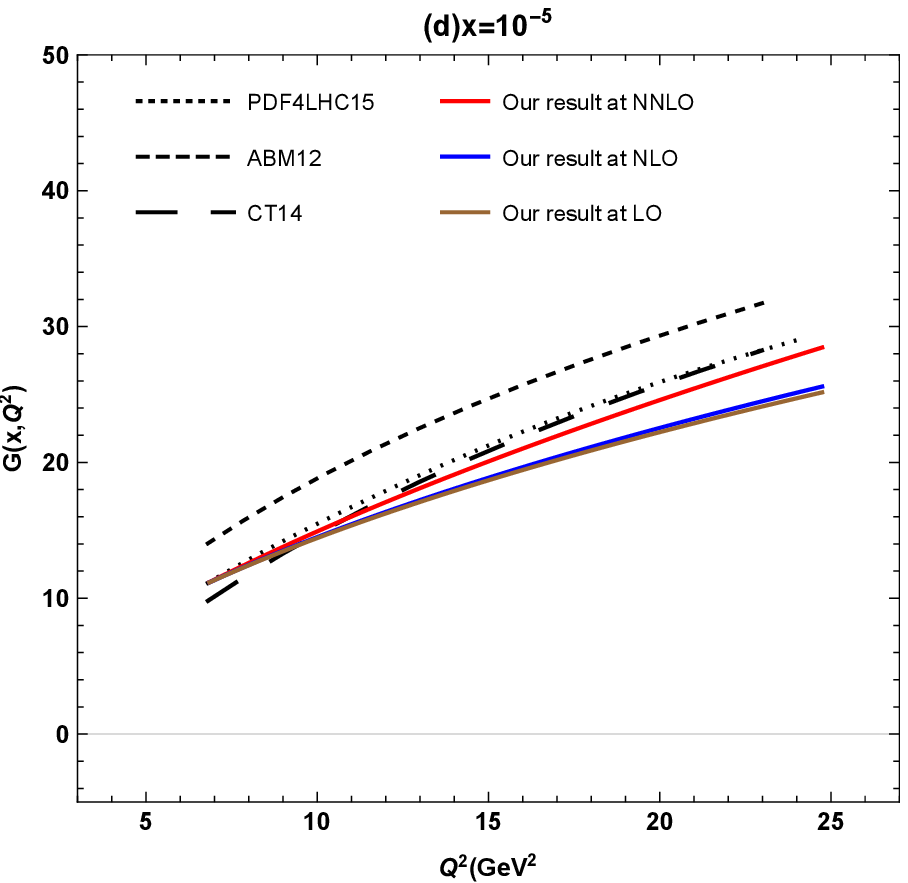}}
		
		\caption{$Q^2$ evolution of $G(x,Q^2)$ for $R=5 GeV^{-1}$. \textit{Solid red lines} are our NNLO result while \textit{solid blue lines} and \textit{solid brown lines} are our NLO and LO results respectively. \textit{Dotted lines} are from the PDF4LHC15 set, \textit{dot dashed lines} are from HERAPDF15 set, \textit{dashed lines} are from NNPDF30 and \textit{absolute dashed lines} are from CT14.}
		
	\end{figure} 
	\\ In this work we have considered the kinematic region to be 6.5 \(\text{GeV}^2\) $\leq $ \(Q^2\) $\leq 25 $ \(\text{GeV}^2\) where all
	our assumptions look natural and our solution seems to be valid. Fig. 1(a-b) shows the plot of \(T^2(t)\) and \(T_0T(t)\) as well as \(T^3(t)\) and \(T_1T(t)\)
	with respect to \(Q^2\). In the range 6.5 \(\text{GeV}^2\) $\leq $ \(Q^2\) $\leq 25 $ \(\text{GeV}^2\), it is observed that for \(T_0=0.0338\) and \(T_1=0.00115\) the difference
	between \(T^2(t)\) and \(T_0T\) as well as \(T^3(t)\)and \(T_1T\) becomes negligible. Fig. 2(a-d) and fig. 3(a-d) represent our best fit results of t evolution
	of gluon distribution function G(x, \(Q^2\)) for R=2 \(\text{GeV}^{-1}\) and for R=5 \(\text{GeV}^{-1}\) respectively for different values of x, viz. \(10^{-2}, 10^{-3},10^{-4}\) and \(10^{-5}\). We have taken the input distribution \(G\left(x,Q_0^2\right.\)) of a given value of initial \(Q^2\) from PDF4LHC dataset and evolve
	GLR-MQ equation. The input $G(x,Q_0^2)$ is taken at an input value of \(Q_0^2\)$\approx $ 6.76 \(\text{GeV}^2\). We have chosen the input from PDF4LHC15
	set since this set is based on the LHC experimental simulations, the 2015 recommendations \cite{34} of the PDF4LHC 	working 
\begin{figure}[h]
	\centering
	\subfloat{%
		\includegraphics[clip,width=0.38\columnwidth]{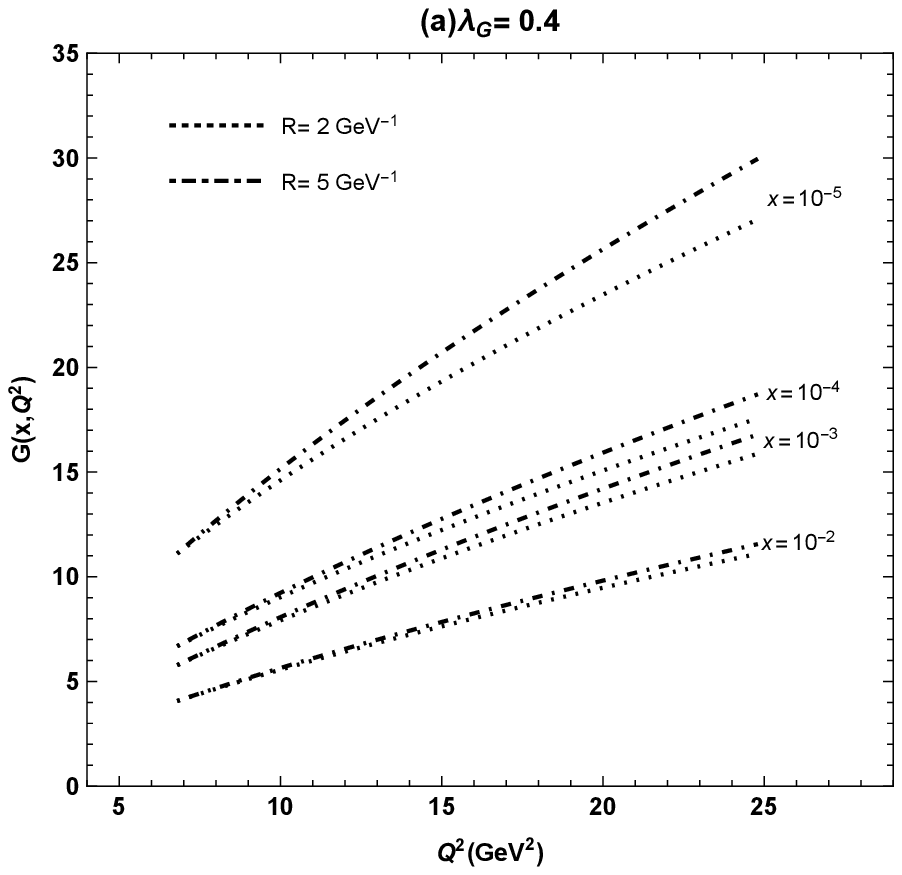}}
	\hspace{5mm}
	\subfloat{%
		\includegraphics[clip,width=0.38\columnwidth]{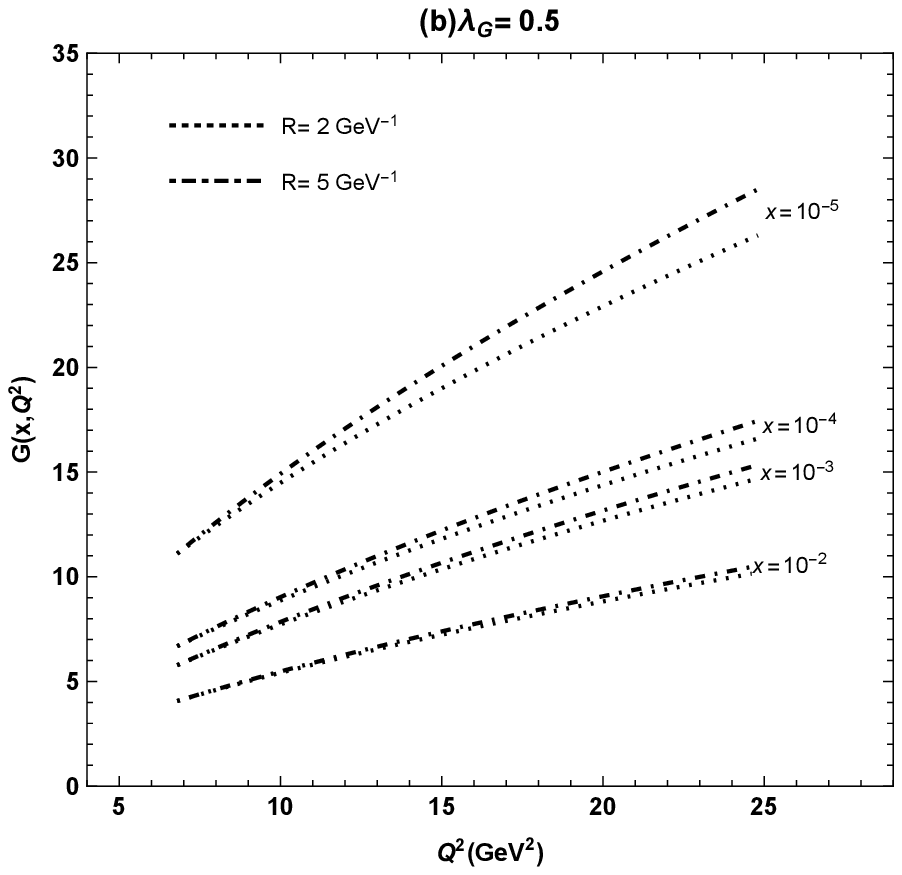}}
	\caption{Sensitivity of $R$ in our results of $Q^2$ evolution of $G(x,Q^2)$.}
	
\end{figure} group and contain combinations
	of more recent CT14, MMHT2014,and NNPDF3.0 PDF ensembles. By the phenomenological analysis we have taken the average value of $\Lambda
	$ to be 0.2 GeV \cite{22,23,24}. In fig. 4(a-b), we have investigated the effect of nonlinearity in our results for two different values of R viz. R=2 \(\text{GeV}^{-1}\) and R=5
	\(\text{GeV}^{-1}\) at different values of x viz. \(10^{-2}, 10^{-3},10^{-4}\) and \(10^{-5}\) respectively. The value of R depends on how the
	gluons are distributed within the proton. If the gluons are distributed over the entire nucleon then R will be of the order of the proton radius
	(R $\simeq $5 \(\text{GeV}^{-1}\)). On the other hand, if gluons are concentrated in hot-spots then R will be very small (R $\simeq $ 2 \(\text{GeV}^{-1}\)) \cite{35}. We have also performed an analysis (fig. 5(a-b)) to check sensitivity of the Regge intercept \(\lambda _G\) in our result by comparing our result of gluon
	distribution \(G\left(x,Q^2\right.\)) for three different values of \(\lambda _G\) viz. 0.4, 0.5 and 0.6 for x = \(10^{-2}, 10^{-3},10^{-4}\) and
	\(10^{-5}\) for both the cases R $\simeq $ 2 \(\text{GeV}^{-1}\) and R $\simeq $ 5 \(\text{GeV}^{-1}\).
	\section{Discussions and conclusion}
	In this work, we have solved the GLR-MQ evolution equation up to next-to-next-leading order (NNLO) by considering Regge like behavior of the gluon distribution
	function. Here we have incorporated the NNLO terms into the gluon-gluon splitting function \(P_{\text{gg}}\)($\omega $) and the running coupling constant
	\(\alpha _s\left(Q^2\right)\). We have examined the validness of Regge behavior of the gluon distribution function in our phenomenologically determined moderate \(Q^2\) kinematic region 6.5 \(\text{GeV}^2\) $\leq $ \(Q^2\)$\leq 25 $ \(\text{GeV}^2\) and \(10^{-5}\) $<$ x $<$ \(10^{-2}\), where nonlinear effects cannot be neglected and it is found that our results show almost similar behavior to those obtained from various global parameterization groups and global fits. We can conclude that our solution is valid only in the vicinity of the saturation border. Our solution of gluon distribution function increases with increase in \(Q^2\) which agree with the perturbative QCD fits at small-x.
	\begin{figure}[b]
		\centering
		\subfloat{%
			\includegraphics[clip,width=0.38\columnwidth]{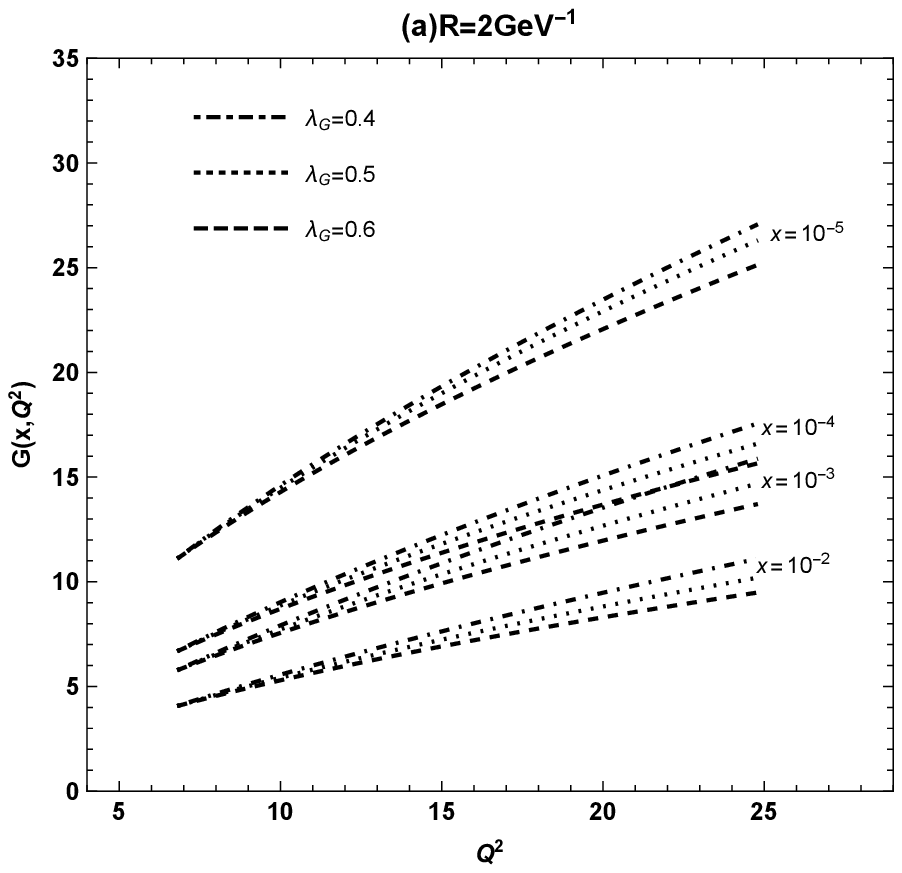}}
		\hspace{5mm}
		\subfloat{%
			\includegraphics[clip,width=0.38\columnwidth]{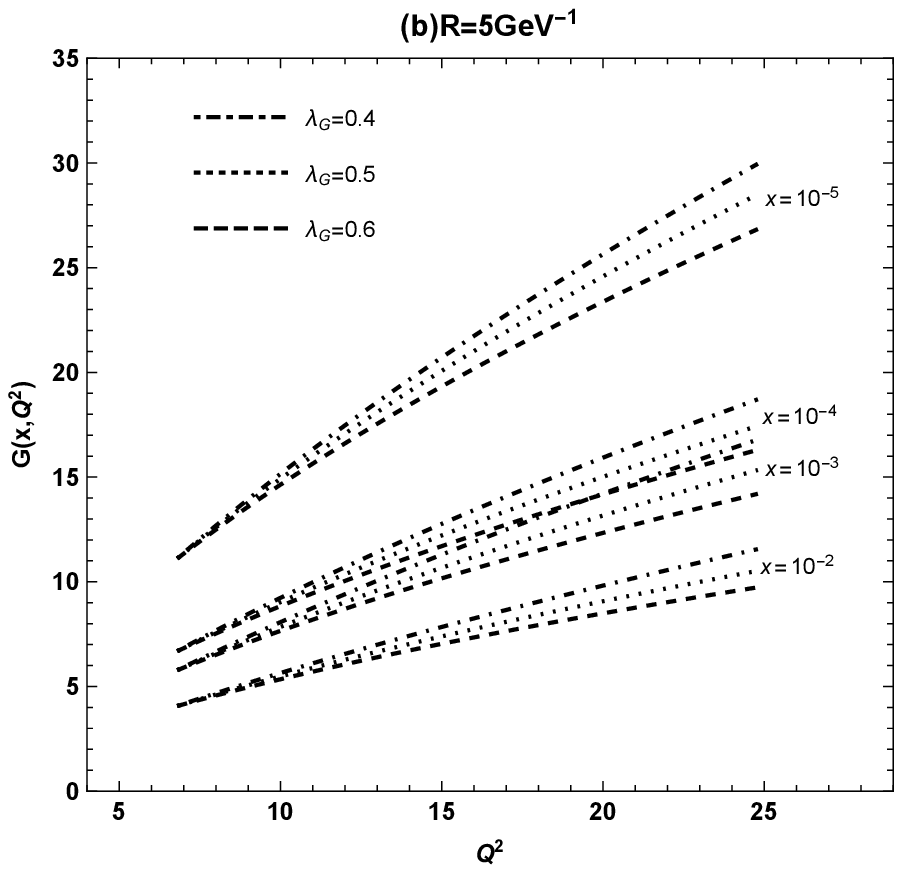}}
		\caption{Sensitivity of $\lambda_G$ in our results of $Q^2$ evolution of $G(x,Q^2)$.}
		
	\end{figure}
	It is also observed that the gluon distribution \(G\left(x,Q^2\right.\)) of our NNLO solution lies slightly above the NLO and LO results as \(Q^2\)	increases and x decreases. This is because of the inclusion of the NNLO terms in the splitting function $P_{gg}(\omega)$ which in fact gives a  better compatibility with different global fits as compared to  NLO and LO results. Again when we go on decreasing x, it is observed that $G(x,Q^2)$ gets more tamed suggesting gluon recombination which appear to be more apparent towards small x. Thus the nonlinear effects are found to play a significant role towards very small x ($\leq $ \(10^{-3}\))  for both cases of  R= 2 \(\text{GeV}^{-1}\) and at R= 5 \(\text{GeV}^{-1}\) as seen in fig. 2(c-d) and fig. 3(c-d). On the other hand nonlinearities vanish rapidly at larger values of x. However it is found that R= 5 \(\text{GeV}^{-1}\) gives better solution towards experimental data than that of R= 2 \(\text{GeV}^{-1}\) at very small x ($\leq $ \(10^{-3}\)). Again as seen in fig. 4 (a-b) the nonlinearity increases with decrease in the values of R however the differences in the results at R= 2 \(\text{GeV}^{-1}\) and at R= 5 \(\text{GeV}^{-1}\) increases with decrease in x. It is found that the gluon distribution function $G(x,Q^2)$
	shows steep behavior at R= 5 \(\text{GeV}^{-1}\) on the other hand taming of $G(x,Q^2)$ is more significant at R= 2 \(\text{GeV}^{-1}\). We have also investigated the effect of nonlinearities in our results for different values of \(\lambda _G\) in fig. 5 (a-b) and found that our solutions are highly sensitive to \(\lambda_G\) towards decrease in x. Finally from this work, we can conclude  that for very small-x ($\leq $ \(10^{-3}\)), our solution of NNLO plays more significant role than that of NLO and LO.
	
	\section*{Acknowledgements}
	One of the authors (M. Lalung) is grateful to CSIR, New Delhi for financial assistantship in the form of Junior Research Fellowship.
\end{document}